\documentclass[12pt]{article}
\usepackage{graphicx}
\begin{document}

\title{Approximative Analytic Study of Fermions in Magnetar's Crust; Ultra-relativistic Plane Waves, Heun and Mathieu Solutions and Beyond}

\vspace{1cm}

\author{Marina--Aura Dariescu and Ciprian Dariescu \\
Faculty of Physics,
^^ ^^ Alexandru Ioan Cuza'' University \\
Bd. Carol I, no. 11, 700506 Ia\c{s}i, Romania\\
Email: marina@uaic.ro}

\date{}
\maketitle

\begin{abstract}
Working with a magnetic field periodic along $Oz$ and decaying in time, we deal with the Dirac-type equation characterizing the fermions evolving in magnetar's crust.
For ultra-relativistic particles, one can employ the perturbative approach, to compute the conserved current density components. 
If the magnetic field is frozen and the magnetar is treated as a stationary object, the fermion's wave function is expressed in terms of the 
Heun's Confluent functions. 
Finally, we are extending some previous investigations on the linearly independent fermionic modes solutions to the Mathieu's equation and we discuss the energy spectrum and the Mathieu Characteristic Exponent.
\end{abstract}

\begin{flushleft}
{\it Keywords}: Magnetars; Heun Confluent functions; Mathieu's equation;  
\\
{\it PACS:} 
02.30.Jr (Partial differential equations);\\
03.65.Pm (Relativistic wave equations);\\
11.15.-q (Gauge field theories); \\
11.10.Lm (Nonlinear or nonlocal theories and models); \\
97.10.Ld (Magnetic and electric fields); \\
97.60.Jd (Neutron stars).
\end{flushleft}

\newpage

\section{Introduction} 

Many years after Duncan and Thompson, [Duncan and Thompson, 1992], introduced the notion of
magnetars, for almost non-rotating neutron stars, whose magnetic field strength was estimated to be about $10^2$ to $10^3$ larger than the one of a neutron star,
this class of astrophysical objects has become an active area of research.

With magnetic fields greater than
the critical induction at which the cyclotron energy of an electron
equals its rest mass energy, they are the only source of the brightest cosmic events originating outside the solar system, known as giant flares.

The magnetars exotic properties have led to many open questions, especially related to
the configuration of the magnetic field inside and to their internal structure
[Sinha and Mukhopadhyay, 2010; Orsaria {\it et al.}, 2011].

Today, it is no doubt that the crust/core coupling [Glampedakis and Andersson, 2006], the 
elastic properties of the outer part of a neutron star
[Pethick and Potekhin, 1998], the
magnetic field geometry and the effect of the strong magnetic pressure on the tension in the star crust, which are all alterating the detected modes frequencies, 
are strongly depending on the crustal composition [Piro and Bildsten, 2006],
and hence on the equation of state (EOS) [Lattimer and Prakash, 2004].

Theoretically, by solving the Einstein equations for a general time-invariant spherically symmetric metric, one comes to the celebrated Tolman--Oppen\-hei\-mer--Volkoff (TOV) equation which is expressing the dependence of the star matter density and pressure on its mass and radius. Reversely, maximum masses and minimum periods of the neutron stars allow us to set constraints on the EOS. 

The present paper is following some previous investigations on
spinless particles, described by the Klein--Gordon equation, moving in
strong magnetic induction periodic along $Oz$ [Dariescu {\it et al.}, 2011a; Dariescu and Dariescu, 2011b].

Since magnetars are being composed of different types of fields which all have deep implications on their parameters [Crawford and Kazanas, 2009], the present work is a general analysis of fermions evolving in the magnetic field configuration proposed by
Wareing and Hollerbach, in their three-dimensional pseudo-spectral numerical MHD code
[Wareing and Hollerbach, 2006; Wareing and Hollerbach, 2009].

\section{Ultra-relativistic Particles in the Crust}

Generally, the covariant theory starts from an effective Lagrangian of the nucleons and as many meson fields as needed to produce the basic nuclear matter. 

By neglecting the self-interaction terms, one comes to the decoupled Dirac and Klein--Gordon-type equations
(in natural units $\hbar = 1 =c$) 
\begin{equation}
\left[ \gamma^i \, D_i + m_0 \right] \Psi \, = \, 0 \, ,
\end{equation}
and 
\begin{equation}
\left[ \eta^{ij} D_i D_j - M^2 \right] \Phi \, = \, 0 \, ,
\end{equation}
where the covariant derivatives,
\[
D_i = \partial_i - iq A_i \; ,
\]
include the interaction with the external fields, $E_y (z,t)$ and $B_x (z,t)$, generated by the $A_y$ component of the four-potential.

By expressing the bi-spinor $\Psi$ in the equation (1) as
\begin{equation}
\Psi = \left[ \gamma^j \, D_j - m_0 \right] \chi \, ,
\end{equation}
and using the Dirac representation for the $\left \lbrace \gamma^i \right \rbrace_{\overline{1,4}}$ matrices
\[
\gamma^{\mu} = -i \beta \, \alpha^{\mu} \; , \; \; \gamma^4 = - i \beta \; , \; \; \mu  = \overline{1, 3} \; ,
\]
with
\[
\beta = \left( \begin{array}{cc} {\cal I} & 0 \\ 0 & - {\cal I}
\end{array} \right)
\; , \; \; \alpha^{\mu} = \left(
\begin{array}{cc}
0 & \sigma^{\mu} \\
\sigma^{\mu} & 0
\end{array}
\right) ,
\]
where ${\cal I}$ is the unit 2x2 matrix and $\sigma^{\mu}$ are the usual Pauli matrices,
the equation (1) leads to [Dariescu and Dariescu, 2012]
[M. A. Dariescu and C. Dariescu, submitted to Astroparticle Physics]
\begin{equation}
\left[ \partial_x^2 + \left( \partial_y - i q A_y \right)^2 +
\partial_z^2 - \partial_t^2 - m_0^2 + q  \Gamma^1 \, B_x  \right]
\chi = 0 \, .
\end{equation}
This contains the matrix $\Gamma_1 \equiv i \gamma^3 \gamma^2$, whose
eigenvectors
\begin{equation}
w_{1,2} = \frac{1}{2} \left[
\begin{array}{r} 1 \\ \pm 1 \\ 1 \\ \pm 1
\end{array}
\right] \; , 
\end{equation}
are corresponding to the eigenvalues $\lambda_{1,2} = \pm 1$.

The explicit form of the equation (4) is depending on the configuration of the background fields in the magnetar's crust. This is a very active field of research and the expression of the strong magnetic induction is not unique.
One may check easily that the following expressions 
\begin{eqnarray}
& &
E_y \, = \, \frac{\kappa}{\sigma} \, b_0 
\sin \left( \kappa z \right) \exp \left[ - \frac{\kappa^2}{\sigma} \, t \right]  , \nonumber \\*
& &
B_x \, = \, b_0  
\cos \left( \kappa z \right) \exp \left[ - \frac{\kappa^2}{\sigma} \, t \right] ,
\end{eqnarray}
sustained by only one component of the four-potential, namely
\begin{equation}
A_y = - \, \frac{b_0}{\kappa} 
\sin \left( \kappa z \right) e^{- \alpha t } \; \; {\rm with} \;
\alpha \equiv \frac{\kappa^2}{\sigma} 
\end{equation}
are solutions of the Maxwell's Equations
\[
\nabla \times \vec{E} = - \, \frac{\partial \vec{B}}{\partial t} \; , \; \;
\nabla \cdot \vec{E} = 0 \; , \; \; \nabla \cdot \vec{B} = 0 \; , \; \;
\nabla \times \vec{B} = \vec{j} +  \frac{\partial \vec{E}}{\partial t} \, .
\]
For a magnetic field frozen in the crust, the time-independent part of the magnetic induction in (6) is of the form proposed by Wareing and Hollerbach
[Wareing and Hollerbach, 2006; Wareing and Hollerbach, 2009], i.e. $B_x = B_0 h(z)$, where $h(z)$ is a periodic function.

With the standard variables separation
\begin{equation}
\chi_{1,2} = \eta_{1,2} (z,t) \exp \left[ i \vec{p}_{\perp} \cdot \vec{x}_{\perp} \right] w_{1,2}
\; ,
\end{equation}
where $\vec{p}_{\perp} \cdot \vec{x}_{\perp} = p_x x + p_y y$
and the index ^^ ^^ 1'' or ^^ ^^ 2'' stands for $\lambda_1 = +1$ and $\lambda_2 = -1$, the equation (4) leads to
\begin{eqnarray}
\frac{\partial^2 \eta}{\partial z^2} - \frac{\partial^2 \eta}{\partial t^2} - \left[ p_{\perp}^2 + m_0^2 + 2 p_y \frac{qb_0}{\kappa} \sin ( \kappa z) e^{- 
\alpha t} + \left( \frac{qb_0}{\kappa} \right)^2 \sin^2 ( \kappa z ) e^{-2 \alpha t} \right] \eta \nonumber \\*
\pm q b_0 \cos ( \kappa z) e^{-\alpha t} 
\eta \, = \, 0 \;  .
\end{eqnarray}

Analytically solving of the above equation in the
general case when all the terms are taken into account is a difficult task and therefore we need some simplifying assumptions.

For ultra-relativistic particles with the momentum along the electric field, $p_y \gg q b_0 / \kappa$, so that $\left( p_y - q A_y \right)^2 \approx p_y^2$,
the equation (9) simplifies to
\begin{equation}
\frac{\partial^2 \eta}{\partial z^2} - \frac{\partial^2 \eta}{\partial t^2} - \left[ p_{\perp}^2 + m_0^2 \right] \eta \pm q b_0 \cos ( \kappa z) e^{-\alpha t}
\eta \, = \, 0 
\end{equation}
and one may use the Laplace--Fourier expansion
\begin{equation}
\eta (z , t ) \, = \, \sum_{n=0}^{\infty} \psi_n (z) e^{-(n \alpha + i \omega )t} \, ,
\end{equation}
to get the following recurrent differential system for the amplitude functions $\psi_n (z)$, 
\begin{equation}
\frac{d^2 \psi_n}{dz^2} + \left[ p^2 - 2 n i \alpha \omega - n^2 \alpha^2 \right] \psi_n \, \pm q b_0 \cos ( \kappa z ) \, \psi_{n-1} \, = 0 \; ,
\end{equation}
where $p^2 \equiv \omega^2 - p_{\perp}^2 - m_0^2$. 

For $n=1$, the relation (12) becomes the inhomogeneous second order linear differential equation with constant
coefficients,
\begin{equation}
\frac{d^2 \psi}{dz^2} + \left[ p^2 - 2 i \alpha \omega - \alpha^2 \right] \psi \, \pm q b_0 \cos ( \kappa z ) \, \psi_0 \, = 0 \, ,
\end{equation}
where
\begin{equation}
\psi_0 = {\cal N} e^{ipz} 
\end{equation}
is satisfying the initial differential equation
\begin{equation}
\frac{d^2 \psi_0}{dz^2} + p^2 \psi_0 \, = \, 0 \, .
\end{equation}
Using in (13) the standard decomposition
\begin{equation}
\psi \, = \, A \, e^{i(\kappa + p ) z} \, + \,
B \, e^{- i(\kappa - p ) z} \, ,
\end{equation}
we are identifying the coefficients,
\begin{eqnarray}
& & A \, = \, \pm \, {\cal N} \, \frac{qb_0}{2} \,  
\frac{\kappa^2 + 2 \kappa p + \alpha^2 - 2 i \alpha \omega}{(\kappa^2+2 \kappa p + \alpha^2)^2+ 4 ( \alpha \omega)^2} \approx 
\pm \, {\cal N} \, \frac{qb_0}{4} \,  
\frac{\kappa p -  i \alpha \omega}{(\kappa p)^2 + ( \alpha \omega)^2} , 
\nonumber \\*
& & B \, = \, \pm \, {\cal N} \, \frac{qb_0}{2} \,  
\frac{\kappa^2 - 2 \kappa p + \alpha^2 - 2 i \alpha \omega}{(\kappa^2-2 \kappa p + \alpha^2)^2+ 4 ( \alpha \omega)^2} \approx 
\pm \, {\cal N} \, \frac{qb_0}{4} \,  
\frac{- \kappa p -  i \alpha \omega}{(\kappa p)^2 + ( \alpha \omega)^2} ,
\nonumber 
\end{eqnarray}
and obtain the following mode expressions
\begin{eqnarray}
\psi_{1,2} (z) & = &  \pm \, {\cal N} \, \frac{qb_0}{2} \,  
\frac{\kappa p \sin ( \kappa z ) - \alpha \omega  \cos ( \kappa z)}{(\kappa p)^2 + ( \alpha \omega)^2} 
\, e^{ip z} \nonumber \\*
& = & \pm \, \frac{{\cal N}}{2} \, \frac{qb_0}{\sqrt{(\kappa p)^2 + ( \alpha \omega)^2}} \, 
\sin \left( \kappa z - \theta \right) e^{ip z} \, ,
\end{eqnarray}
where
\[
\tan \theta = \frac{\alpha \omega}{\kappa p } \, ,
\]
is a phase factor in the wave functions of chiral ultra-relativistic fermions of energy $\omega$, evolving in the magnetic field $B_x = b_0 e^{- \alpha t} \cos \left( \kappa z \right)$. 

Going all the way back, the two linearly independent modes of positive energy, solutions of (1), are 
\begin{equation}
\Psi_1  \, = \, 
\frac{1}{2} \left[ 
\begin{array}{c}
p_x - i p_y + p - \omega + i \alpha - m_0  - i h(z)  \\ \\
p_x + i p_y -p  - \omega + i \alpha - m_0 +
i h(z) \\ \\
- p_x + i p_y -p  + \omega - i \alpha - m_0 + i h(z) \\ \\
- p_x - i  p_y +p + \omega - i \alpha - m_0 - i h(z)
\end{array}
\right] \, \psi_1 e^{\left[ i \left( \vec{p}_{\perp} \cdot \vec{x}_{\perp} - \omega t \right) \right]} \,  e^{- \alpha t} 
\end{equation}
and
\begin{equation}
\Psi_2 \, = \, \frac{1}{2}
\left[ 
\begin{array}{c}
-p_x + i p_y  +p - \omega + i \alpha - m_0  - i h(z)  \\ \\
p_x + i p_y +p + \omega - i \alpha - m_0 -
i h(z)
\\ \\
p_x - i p_y -p + \omega - i \alpha - m_0 + i h(z) \\ \\
- p_x - i p_y -p- \omega + i \alpha - m_0 + i h(z) 
\end{array}
\right] \psi_2
\, e^{\left[ i \left( \vec{p}_{\perp} \cdot \vec{x}_{\perp} - \omega t \right) \right]} \,  e^{- \alpha t} \, , \nonumber \\* 
\end{equation}
where we have introduced the notation
\[
h(z) = \, \kappa \frac{\kappa p \cos ( \kappa z) + \alpha \omega \sin( \kappa z)}{\kappa p \sin ( \kappa z) - \alpha \omega \cos ( \kappa z)} \, =
\kappa \, \cot \left( \kappa z - \theta \right) .
\]

For the superposition
\begin{equation}
\Psi = \frac{1}{2} \left( \Psi_1 + \Psi_2 \right) = 
\frac{1}{2} \left[ 
\begin{array}{c}
p_x - i p_y  \\ \\
-p  - \omega + i \alpha  +
i h(z) \\ \\
- p_x + i p_y  \\ \\
p + \omega - i \alpha  - i h(z)
\end{array}
\right] \, \psi_1 e^{\left[ i \left( \vec{p}_{\perp} \cdot \vec{x}_{\perp} - \omega t \right) \right]} \,  e^{- \alpha t} \, , 
\end{equation}
the current density components defined by
\begin{equation}
\vec{j}  = q \Psi^{\dagger} \vec{\alpha} \Psi \; \; , \; \;
j^4 = Q = q \Psi^{\dagger} \Psi \, ,
\end{equation}
have the explicit expressions
\begin{eqnarray}
j_x & = & q \left[ \left( \omega + p \right) p_x + \left( \alpha + h \right) p_y  \right] | \psi_1 |^2 \,
e^{-2 \alpha t} \nonumber \\*
j_y & = & 
q \left[ \left( \omega + p \right) p_y - \left( \alpha + h \right) p_x  \right] | \psi_1 |^2 \,
e^{-2 \alpha t}
\nonumber \\*
j_z & = &
\frac{q}{2} \left[ 2 p^2 + 2 \omega p + m_0^2 + \left( \alpha + h \right)^2 \right] | \psi_1 |^2 \, e^{-2 \alpha t} \, 
\nonumber \\*
Q & = &
\frac{q}{2}  \left[ 2 \omega^2+ 2 \omega p - m_0^2 + \left( \alpha + h \right)^2 \right] | \psi_1 |^2 \, e^{-2 \alpha t} \, ,
\end{eqnarray}
which are fulfilling the conservation law $\partial_i j^i =0$, for $\alpha =0$ and the resonance condition
\begin{equation}
\kappa^2 = 2 p^2 + 2 \omega p + m_0^2 \, .
\end{equation}

For $\alpha \neq 0$, at $z = L_z$, once the longitudinal momentum characterizing 
the particles confined in the crust is equal to 
\[
p = \frac{\alpha \omega}{\kappa} \cot \left( \kappa L_z \right) \, ,
\]
i.e. $\kappa L_z = \theta$, the wave function absolute value and consequently
$j_x$ and $j_y$ are all vanishing, while the other current's components, at $z=L_z$, i.e.
\[
j_z (z=L_z) = Q (z=L_z) = \frac{q}{2} \left| {\cal N} \right|^2 \left( \frac{qb_0}{2} \right)^2 \frac{\kappa^2}{( \alpha \omega )^2} \sin^2 \left( \kappa L_z \right)
e^{-2 \alpha t} \; ,
\]
are satisfying $\partial_z j_z + \partial_t Q =0$. 

\section{HeunC Solution within Ohmic Timescale}

In the opposite case, corresponding to $p_y \ll q b_0 / \kappa$ and
the time variable less than the {\it characteristic time}, so that $e^{- \alpha t} \approx 1$,
which is a more realistic situation for a magnetar characterized by a strong magnetic induction, $b_0 \sim 10^{15}$ G,
the four-potential component (7) can be taken as
\begin{equation}
A_y = - \, \frac{b_0}{\kappa} \, \sin \kappa z \, .
\end{equation}
With the function $\eta (z ,t)$, in the general equation (9), written as
\[
\eta (z,t) \, = \varphi (z) e^{-i \omega t} \, ,
\]
the chiral components $\varphi_{1,2} (z)$, corresponding to $\lambda = \pm 1$, 
are solutions of the equations
\begin{equation}
\frac{d^2 \varphi_{1,2}}{d \zeta^2} + \left[ \frac{p^2}{\kappa^2} - \frac{b^2}{\kappa^4}  \sin^2 \zeta \pm \frac{b}{\kappa^2} \cos \zeta \right] \varphi_{1,2} \, = \, 0 \, ,
\end{equation}
where $\zeta \equiv \kappa z$ and $q b_0 \equiv b$.

In terms of the new variable
$s = \cos \zeta$, that maps the $\zeta$ space to $s \in \left( -1 \, , \, 1 \right)$, with
\begin{equation}
\varphi_{1,2} (s) \, = \exp \left[ \pm \frac{b}{\kappa^2} s \right] u_{1,2}(s) \, ,
\end{equation}
in (25), we come to the following differential equation for $u$,
\begin{equation}
(1-s^2) \frac{d^2u_{1,2}}{d s^2} + \left[ \pm 2 \frac{b}{\kappa^2} \left(1-s^2 \right) - s \right] \frac{d u_{1,2}}{d s} + \frac{p^2}{\kappa^2} \, u_{1,2} = 0
\end{equation}
Up to normalization constants, the solutions of (27) are the $HeunC$ confluent functions
[Arscott, 1995] 
\begin{eqnarray}
u_{1,2} ( s ) & \in & \left \lbrace HeunC \left[ \pm 4 \frac{b}{\kappa^2} , \, - \frac{1}{2} , \, - \frac{1}{2} , \, \mp 2 \frac{b}{\kappa^2} , \, \frac{3}{8} - \frac{p^2}{\kappa^2} \pm \frac{b}{\kappa^2} , \, \frac{s+1}{2} \right] , \right.
\nonumber \\*
& & \left.  \sqrt{2(1+s)} HeunC \left[ \pm 4 \frac{b}{\kappa^2} , \, \frac{1}{2} , \, - \frac{1}{2} , \, \mp 2 \frac{b}{\kappa^2} , \, \frac{3}{8} - \frac{p^2}{\kappa^2} \pm \frac{b}{\kappa^2} , \, \frac{s+1}{2} \right] \right \rbrace ,
\nonumber \\*
\end{eqnarray}
with the boundary conditions
$u(s = \pm 1) =0$ and $u^{\prime} ( s= \pm 1) = finite$, so that the wave functions (26) have the explicit expressions
\begin{eqnarray}
\varphi_{1,2} ( \zeta ) & = & \exp \left[ \pm \frac{b}{\kappa^2} \cos \zeta
\right] u_{1,2} \left( \cos^2 \frac{\zeta}{2} \right) ,
\end{eqnarray}
where $u_{1,2}$ are given in (28).

Now, one may follow the procedure developed in [Gurappa and Panigrahi, 2004], for a differential equation which can be cast in the form
\[
\left[ F(D) + P(x , \, d/dx ) \right] u(x) = 0 \, ,
\]
where $D \equiv x \frac{d \;}{dx}$, $F(D) = \sum_n a_n D^n$ is a diagonal operator in the space of monomials and $P(x , \, d/dx )$ is an arbitrary polynomial function of $x$ and $d/dx$. The necessary condition for a polynomial form of $u$ is
\[
F(D) x^n = 0 \, ,
\]
leading, for
\[
F(D) = - \, D^2 - \frac{b^2}{\kappa^4} + \frac{p^2}{\kappa^2}
\, ,
\]
in (27), to the
following longitudinal momentum quantization law
\[
p^2 = n^2 \kappa^2 + \frac{b^2}{\kappa^2} \, ,
\]
or, for $b \ll \kappa$, to the familiar resonance relation
$p = n \kappa$.

Even though we have derived an analytic solution to the Dirac equation (1) for the fermions minimally coupled to the electromagnetic field generated by the potential component (24), 
we must signal the fact that there are unsolved problems regarding the normalization of the Heun functions and their behavior for some domains of the complex variable.
In terms of available soft, their treatment can be done only with MAPLE routines which are working well in particular cases, but are very sensitive to the parameters, breaking down for values which might be of interest. 

Finally, if we keep only the first order contribution in $b/\kappa^2$,
the equation (25) becomes the well known Mathieu's equation
[Gradshteyn and Ryzhik, 1965]
\begin{equation}
\frac{d^2 y}{d w^2} + \left[ \alpha - 2 \beta \, \cos ( 2 w ) \right] y
 \, = \, 0 \, ,
\end{equation}
with $w = \kappa z / 2$ and
\[
\alpha = \frac{4p^2}{\kappa^2} \; , \; \; \beta = \mp \frac{2b}{\kappa^2} \, ,
\]
whose solutions are the so-called Mathieu's even and odd functions ${\cal C}$ and ${\cal S}$,
\begin{equation}
\varphi_{1,2} (z) \, = \, \left \lbrace {\cal C} \left[ \frac{4p^2}{\kappa^2} \, , \, \mp \frac{2b}{\kappa^2} \, , \, \frac{\kappa z}{2} \right] , \;
{\cal S} \left[ \frac{4p^2}{\kappa^2} \, , \, \mp \frac{2b}{\kappa^2} \, , \, \frac{\kappa z}{2} \right] \right \rbrace \, .
\end{equation}

These can be written as
\[
\varphi \sim e^{i \gamma w} f(w) \, ,
\]
where $f$ is a periodic function and the so-called Mathieu Characteristic Exponent (MCE), $\gamma$, is strongly related to the
particles momentum along $Oz$ and to the external magnetic field configuration. 

For $\alpha \gg \beta$, the solution of (31) is stable and the energy spectrum remains continuous almost everywhere. In our case, this means ultra relativistic particles with high momentum along $Oz$, i.e.
\[
p^2 c^2 = \omega^2 - p_{\perp}^2 c^2 - m_0^2 c^4  \gg \frac{\hbar q b_0 c^2}{2} \, ,
\]
where $\hbar q b_0 c^2 \approx 4 \; ( MeV)^2$, for $b_0 \sim 10^{15} \; G$, which is typical for the magnetic induction in magnetar's crust. 

Reversely, for fermions characterized by $\beta \gg \alpha$, the periodic term in (31) can no longer be treated as a perturbation, and one gets discrete energy eigenvalues, separated by large instability regions, which become wider as $\beta$ increases. The asymptotic expansion [Beckert {\it et al}, 1979]
\begin{equation}
\alpha \approx - 2 \beta + 4n \sqrt{\beta} \, , 
\end{equation}
leads to the following energy quantized levels
\begin{equation}
\omega^2 \approx p_{\perp}^2 + m_0^2 + n\kappa \sqrt{2qb_0} \pm q b_0 \, . 
\end{equation}

Moreover, for particles whose longitudinal momentum is close to
$\kappa/2$, in the range
\begin{equation}
\frac{\kappa}{2} - \frac{b}{2 \kappa} < p < \frac{\kappa}{2} + \frac{b}{2 \kappa}
 \, ,
\end{equation}
the imaginary part of MCE comes into play, leading to exponentially growing wave functions.

This can be seen by writing down the solution of the general Mathieu's equation (31)
as
\[
y = \sum_{n} C_{2n} e^{i(\gamma + 2n) w} \, ,
\]
where the $C-$coefficients are satisfying the relation 
\[
\alpha C_{2n} - ( \gamma +2n)^2 C_{2n} - \beta C_{2(n-1)} - \beta C_{2(n+1)} \,
\equiv \, 0 \, , 
\]
or, in our case,
\[
C_{2n} + \xi_{2n} \left[ C_{2(n+1)} + C_{2(n-1)} \right] \, = \, 0 \; , \; \;
{\rm with} \; \xi_{2n} = \pm \, \frac{2b}{4 p^2 - (\gamma+2n)^2 \kappa^2}  \, .
\]
By imposing the vanishing of the determinant associated to
the above homogeneous system of unknown $C_{-2(n+1)} , \dots , C_{2(n+1)}$,
we get the following relation for the MCE
\begin{equation}
\cos ( \pi \gamma ) = 1 - 2 \Delta (0) \sin^2 \left[ \pi \frac{p}{\kappa} \right] ,
\end{equation}
where $\Delta (0)$ is the value determinant
if we set $\gamma = 0$ and $p \neq n \kappa$.
Generally, $\gamma$ is real or imaginary, depending on the model parameters. 

For $p$ in the range (35), i.e.
\[
\frac{\pi}{2} \left[ 1 - \frac{b}{\kappa} \right] < \pi \frac{p}{\kappa} <
\frac{\pi}{2} \left[ 1 + \frac{b}{\kappa} \right] , \; \; b \ll  \kappa \; ,
\]
as soon as $\gamma$ becomes imaginary and $\left| \cosh ( \gamma \pi ) \right| >1$, the wave function is not bounded on the real axis [Grib {\it et al.}, 1994], while
for the resonance condition $p = n \kappa$, the relation (36) is replaced with
[Coisson {\it et al.}, 2009]  
\[
\cos ( \pi \gamma ) = 2 \Delta (1) - 1 \; .
\]

For bosons, we have come to similar results [Dariescu and Dariescu, 2011], they being described by the wave function
\[
\Phi = e^{i \left( \vec{P}_{\perp} \cdot \vec{x}_{\perp} - \Omega t \right)} f (z)
\]
where
\[
f(z) \, = \, \left \lbrace {\cal C} \left[ \frac{P_z^2}{\kappa^2} - \frac{b^2}{2 \kappa^4} \, , \, \frac{b^2}{4 \kappa^4} \, , \, \kappa z \right] ,
{\cal S} \left[ \frac{P_z^2}{\kappa^2} - \frac{b^2}{2 \kappa^4} \, , \, \frac{b^2}{4 \kappa^4} \, , \, \kappa z \right] \right \rbrace ,
\]
and by the energy quantization law
\begin{equation}
\Omega^2 \approx P_{\perp}^2 + M^2 + 2 n \, q b_0 \; .
\end{equation}

\section{Conclusions}

As in our previous works, [Dariescu {\it et al.}, 2011a; Dariescu and Dariescu, 2011b], we have considered that the periodic magnetic field proposed by
Wareing and Hollerbach [Wareing and Hollerbach, 2006] is likely to exist in magnetar's crust and we have analyzed how the time and spatial distribution of the external fields (6) have an influence on the wave function, solutions to the corresponding Dirac equation.

In a perturbative approach, for ultra-relativistic particles, we have derived the current density components and the charge density (22). These can be used for studying the back-reaction of the new fields
generated by the Maxwell equations.

Within the Ohmic timescale, 
when the magnetic field is frozen in the crust and the magnetar is treated as a stationary object, the fermions wave functions can be written in terms of the 
Heun's Confluent functions. 

Today, different types of Heun's functions are considered as successors of the
hypergeometric functions, with a wide application in modern physics and 
new algorithms have been created,
for finding solutions to systems of nonlinear transcendental equations
[Fiziev and Staicova, 2012].
For example, the confluent Heun functions have been worked out in the context of the
quasinormal modes for nonrotating black holes
[Fiziev and Staicova, 2011]
or as solutions to Schr\"{o}dinger equation, for different rational
potential functions, in thick braneworlds
[Cunha and Christiansen, 2011].

The final part of the present paper, which extends some previous works [Dariescu {\it et al.}, 2011a; Dariescu and Dariescu, 2011b], is leading to conclusions in agreement with the linear stability approach, belonging to Rheinhardt and Geppert [Rheinhardt and Geppert, 2001], who have proved that the transfer of magnetic energy from the background field to small-scale modes may produce Hall instabilities, at some ranges of wavenumber values.

As we have mentioned previously, the configuration of the background magnetic field in the magnetar's crust has been extensively investigated and is not unique. For example, for a magnetic field oriented along $Ox$ and everywhere
parallel to the surface of a slab with a finite
$z-$thickness, the induction has been written as
$B_x = B_0 \, f(z)$, where $f(z)$ is vanishing for $z > L$.
Guided
by the boundary conditions and working with dimensionless variable $z/L$, the authors of 
[Rheinhardt and Geppert, 2002]
have proposed the form
$f = \left(1 - \frac{z^2}{L^2} \right)$.
With this choice, the ultra-relativistic particles would be described by the differential equation  
\begin{equation}
\frac{d^2 \psi}{dz^2} + \left \lbrace p^2 \pm b \left[ 1 - \left( \frac{z}{L} \right)^2 \right] \right \rbrace \psi = 0 \, ,
\end{equation}
where $b = q B_0$ and $p^2 = \omega^2 - p_{\perp}^2 - m_0^2$.
This is a particular case of the analysis developed in the present work since
one can easily come to the above equation by
performing, in (10) written for $\alpha \approx 0$ and $\eta = \psi (z) e^{-i \omega t}$, the series expansion
\[
\cos ( \kappa z ) \approx 1 - \frac{\kappa^2 z^2}{2} \, ,
\]
valid for $z \ll L$ and by making the identification $\sqrt{2} / \kappa \to L$.

As a general remark, we would like to mention that 
the present work could be extended in several directions. 
 
Thus, because besides leptons, in  dense nuclear matter, the nucleons interact among each other through three meson fields: the isoscalar-scalar meson $\sigma$, isoscalar vector meson $\omega$ and isovector-vector meson $\rho$, one may assume that
currents and charge density produced by the single-particle spinors, as the ones in (22), are actually acting as sources for the Klein-Gordon 
Gordon equations for the time- and space-like meson
fields
\begin{eqnarray}
& &
\left( \Delta - m_{\phi}^2 \right) \tilde{\phi} = g_{\phi} j^p 
\nonumber \\*
& & 
\left( \Delta - m_{\phi}^2 \right) \tilde{\phi}_0 = g_{\phi} Q^p 
\end{eqnarray}
where $\tilde{\phi} = \left \lbrace \vec{\omega} \, , \, \vec{\rho} \right \rbrace$ and $\phi_0 = \left \lbrace \omega_0 \, , \, \rho_0 \right \rbrace$, $g_{\phi}$ are
the respective coupling constants, $m_{\phi}$ are the meson masses
and (see (21))
\[
\vec{j}^{\; p} = q \sum_{i=1}^Z 
\Psi^{\dagger}_i \vec{\alpha} \, \Psi_i \; \; , \; \;
Q^p = q \sum_{i=1}^Z \Psi^{\dagger}_i \Psi 
\]
for a number of protons $Z$.

In the Hartree approximation and considering only static configurations, the system (39)
is producing new meson fields
and one should look for a closed set of solutions which usually is found iteratively, until convergence results. For a detailed study of the effect of strong magnetic field on nuclei in the crust of magnetars and its consequences on the physical measurable quantities, we recommend [Artega {\it et al.}, 2011].

Last but not least, the relativistic particles energy spectrum can be employed in dealing with the Equation of State and its implications on the detected modes frequencies.
This possible extension is very challenging due to the
intense activity on several types of EOS, within different theoretical techniques and for different compositions of the neutron star, which
can account for the observed frequencies and magnetic field strengths.

\end{document}